# Quantum micro-nano devices fabricated in diamond by femtosecond laser and ion irradiation

*Shane M. Eaton[†], J. P. Hadden[†], Vibhav Bharadwaj[†], Jacopo Forneris, Federico Picollo, Federico Bosia, Belen Sotillo, Argyro N. Giakoumaki, Ottavia Jedrkiewicz, Andrea Chiappini, Maurizio Ferrari, Roberto Osellame, Paul E. Barclay, Paolo Olivero, Roberta Ramponi\**

Dr. S.M. Eaton, Dr. V. Bharadwaj, Dr. B. Sotillo, A.N. Giakoumaki, Dr. R. Osellame, Prof. R. Ramponi
Istituto di Fotonica e Nanotecnologie-Consiglio Nazionale delle Ricerche (IFN-CNR) and Dipartimento di Fisica, Politecnico di Milano, Piazza Leonardo da Vinci 32, Milano 20133, Italy
E-mail: roberta.ramponi@polimi.it

Prof. P. E. Barclay
Institute for Quantum Science and Technology, University of Calgary, Calgary T2N 1N4, Canada.

Dr. J. P. Hadden
School of Physics and Astronomy, Cardiff University, Cardiff CF24 3AA, UK

Dr. F. Picollo, Dr. F. Bosia, Prof. P. Olivero
Department of Physics and "NIS" inter-departmental centre, University of Torino, via P. Giuria 1, Torino 10125, Italy

Dr. J. Forneris
Istituto Nazionale di Fisica Nucleare, Sezione di Torino, via P. Giuria 1, Torino 10125, Italy

Dr. O. Jedrkiewicz
Istituto di Fotonica e Nanotecnologie, CNR, Università dell'Insubria Department of Science and High Technology, Via Valleggio 11, Como 22100, Italy

Dr. A. Chiappini, Prof. M. Ferrari
Istituto di Fotonica e Nanotecnologie-Consiglio Nazionale delle Ricerche (IFN-CNR), Characterization and Development of Materials for Photonics and Optoelectronics (CSMFO) and The Centre for Materials and Microsystems (FBK-CMM), Trento 38123, Italy

[†]     These authors contributed equally to the work.

*corresponding author: roberta.ramponi@polimi.it

Diamond has attracted great interest as a quantum technology platform thanks to its optically active nitrogen vacancy (NV) center. The NV's ground state spin can be read out optically exhibiting long spin coherence times of ~1 ms even at ambient temperatures. In addition, the energy levels of the NV are sensitive to external fields. These properties make NVs attractive as a scalable platform for efficient nanoscale resolution sensing based on electron spins and for quantum information systems. Diamond photonics enhances optical interaction with NVs, beneficial for both quantum sensing and information. Diamond is also compelling for microfluidic applications due to its outstanding biocompatibility, with sensing functionality provided by NVs. However, it remains a significant challenge to fabricate photonics, NVs and microfluidics in diamond.

In this Progress Report, an overview is provided of ion irradiation and femtosecond laser writing, two promising fabrication methods for diamond based quantum technological devices. The unique capabilities of both techniques are described, and the most important fabrication results of color center, optical waveguide and microfluidics in diamond are reported, with an emphasis on integrated devices aiming towards high performance quantum sensors and quantum information systems of tomorrow.

**1. Introduction**

Diamond has established itself as an important material system for quantum photonics due to the existence of stable and coherent fluorescent and spin active color centers such as the nitrogen vacancy (NV) center which can be utilized as a single photon source and as a solid-state qubit with long spin coherence time allowing for example long distance entanglement[1]. The biocompatibility of diamond[2], and the possibility to use high density ensembles of the same color centers as room temperature quantum enabled sensors[3,4] presents exciting possibilities for life science microscopy and microfluidic applications. However, to harness diamond's strengths it is essential to have the ability to create micro and nano structures such as waveguides to guide and manipulate light to link together distant NV centers; to generate single or high density ensembles of color centers on demand; and to selectively etch micro channels to enable microfluidic lab-on-chip functionalities in diamond.

Diamond exhibits extreme hardness and chemical resistance, and it is difficult to grow thin layers of high-quality single crystal diamond hetero-epitaxially, which means it is not convenient to use traditional sacrificial layer etching techniques to achieve photonic structures. Even though there has been impressive progress using non-conventional etching techniques, the creation of high-quality photonic structures containing NVs which retain their intrinsically high optical and spin coherence remains a challenge.

Recently, femtosecond laser writing and ion irradiation have emerged as alternative techniques for generating photonic and microfluidic structures and for on-demand positioning of NV and of other color centers. In this progress report, we briefly review the NV center as a resource for quantum technologies and sensing, and the state of the art in fabrication of diamond photonics. We then describe how femtosecond laser writing and ion irradiation can be exploited to form photonic devices in diamond. We describe progress in focused ion beam implantation and femtosecond laser writing for the production of buried waveguides and Bragg reflectors in diamond, and the creation of single and ensembles of color centers for quantum and sensing applications. Finally, we describe the creation of buried microfluidic structures through ion implantation and surface microfluidic channels through femtosecond writing using Bessel beams.

## 2. Background

### 2.1 Color centers in diamond

The NV center occurs when two adjacent sites in diamond's carbon lattice is replaced by a nitrogen and a vacancy, as shown in **Figure 1(a)**. They can appear both as single isolated defects, and in high density ensemble. In its negatively charged state, the electronic ground state forms a spin triplet which can be polarized under green excitation, with the resulting fluorescence showing a zero-phonon line (ZPL) at 637 nm, as represented in **Figure 1(b)**. As one of the spin states fluoresces more brightly than the other, optical readout of the spin state is possible[5].

Diamond's largely spin-less carbon lattice enables exceptional room temperature spin coherence of up to ~1 ms, comparable to trapped ions[6]. Since the spin states are sensitive to magnetic[7] and electric fields[8] through the Zeeman and Stark effects, respectively, NVs can be exploited for quantum sensing of electromagnetic fields with remarkable sensitivity and resolution[9]. Depending on the diamond sample's substitution nitrogen density, NV centers can appear both as single isolated defects, and in high density ensembles.

The two main grades of CVD diamond are electronic grade with intrinsic nitrogen concentration of about 5 ppb and optical grade with a nitrogen concentration of about 100 ppb. For applications which require ensembles of NVs, diamond samples made by high pressure-high temperature growth (HPHT, single crystal, Type Ib) with nitrogen concentrations of 100 ppm and above may be employed.

NVs and other quantum emitters may be created during CVD growth through in-situ doping of the crystal with desired gaseous impurities. This approach enables an accurate precision in the depth distribution of dopants[10,11], and offers a strategy to achieve optimal opto-physical properties for the created color centers[6,12]. On the other hand, the high atomic density of the diamond lattice prevents efficient incorporation of heavy impurities with large atomic radii[13,14].

Furthermore, in-situ doping cannot offer control of the spatial position at which individual optical centers are formed.

## 2.2 Overview of diamond nanofabrication methods

Apart from ion irradiation and femtosecond laser writing, a limited number of other techniques have been successfully applied to the nanofabrication in diamond. Hybrid diamond optical waveguides have been formed using a high refractive index material on a diamond substrate[15,16], but are limited by weak evanescent coupling to NVs near the surface. Another approach uses heterogeneous diamond growth followed by oxygen plasma etching to create membranes to fabricate photonic devices[17–19], however it is difficult to remove tens of microns of diamond while maintaining smooth features needed for photonics applications. Suspended triangular nanobeam waveguides were demonstrated using an angled plasma etching process[20], however a custom Faraday cage was required for each optical circuit. Uniform etching has been achieved by engineering the sample holder to rotate and tilt during the etching process[21]. Angle etching techniques have been used to fabricate race-track resonators and nanobeam structures. Meanwhile, a different technique which exploits quasi isotropic etching using a zero-bias $O_2$ plasma to achieve undercutting has been used to form high Q/V microdisks and nano-beam waveguides[22,23].

## 2.3 Ion beam Lithography

The use of ion beams in the keV-MeV energy range for lithography is an appealing technique, with specific advantages over more conventional approaches such as photolithography or electron-beam lithography [24,25]. It has been widely used for large-scale production of doped integrated electronics, and for the modification of structural, electronic and optical properties of materials. Ion irradiation techniques have the potential to achieve a significant impact in Quantum Technologies. At extremely low ion fluxes, the development of new focusing techniques and of highly sensitive particle detection systems allow individual atoms to be implanted into a solid state target with nanometer precision, with enticing opportunities for deterministic doping of semiconductor-based quantum devices[26] and the creation of single optically active centers in wide-bandgap semiconductors[27]. Higher flux ion beams have been employed for innovative fabrication schemes in diamond, overcoming its chemical and physical properties which pose substantial challenges for conventional lithographic techniques[28].

**2.4 Femtosecond laser writing**

The previously-described lithographic and ion beam irradiation methods have enabled impressive diamond NV-based devices[29], but are largely limited to the fabrication of surface planar structures. In addition, the photonic structures are not easily compatible with fiber optics. Overcoming these restrictions, femtosecond laser writing has been recently demonstrated to achieve 3D microfabrication of high-performance quantum technologies in diamond[30,31].

Femtosecond laser writing was first applied to glasses to inscribe optical waveguides[32], and has since been extended to other passive[33] and active glasses[34], polymers[35], crystals[36] and polymerizable liquid crystals[37]. The versatile technique exploits tightly focused femtosecond laser pulses, which results in a localized refractive index change due to nonlinear absorption at the beam focus[38,39]. In most glasses, the laser irradiation drives an increased refractive index relative to the pristine material, and by scanning the sample with respect to the incident beam, 3D photonic networks may be formed (**Figure 2(a)**). This modality of laser waveguide writing is referred to as type I.

In crystals, femtosecond laser pulses generally lead to a decrease in refractive index at the focus. In this case, an alternative strategy known as Type II is typically employed, in which two closely spaced modification lines are laser written, resulting in a stressed central region capable of guiding light (**Figure 2(b)**). By tailoring the geometry of the modification, photonic structures have been realized in crystals such as lithium niobate ($LiNbO_3$)[40], silicon[41], KGW[42], yttrium aluminium garnet (YAG)[43] and zinc selenide (ZnSe)[44]. In section 3.2, we will describe how femtosecond laser writing can be applied to diamond to form high performance photonics.

In addition to the 3D advantage of femtosecond laser processing, the highly versatile method can be applied in other modalities of microfabrication. It can be used for bulk modification followed by chemical etching to form buried microchannels in glasses[45] and YAG[46]. Alternatively, the laser focus may be translated along the surface to form microfluidic channels or other microfeatures[47] in both transparent and absorbing materials.

## 3. Photonic components in diamond

### 3.1 Ion beam fabrication of photonic components

Energetic ion beams offer several avenues for the fabrication of integrated photonic structures in diamond: i) the use of MeV ion beams for the implementation of lift-off schemes that allow the fabrication of free-standing membranes into which 2D photonic structures can be fabricated; ii) the use of scanning keV Focused Ion Beams (FIB) for micro/nano-patterning of photonic structures; iii) the use of ion-beam-induced structural damage to modify the refractive index of diamond, enabling direct writing of waveguides or fine-tuning of cavity modes[48].

The lift-off technique in diamond was developed in the early 1990s as a method to remove thin layers from bulk samples[49,50]. It is based on the employment of MeV ion beams for the creation of sub-superficial damaged layers formed near to the end of range of the ions, the so-called Bragg peak. Upon thermal annealing, the sub-superficial regions damaged beyond a critical level, known as the graphitization threshold[51], are converted to a sacrificial polycrystalline graphitic layer that can be selectively etched using wet chemical etching, electrochemical etching or annealing in oxygen/ozone. This allows the production of free-standing membranes whose thickness is determined by the species and energy of employed ions.

Alternatively, multiple ion implantations at different depths can be employed to fabricate thinner membranes between two contiguous sacrificial layers[52]. Such membranes are even smoother than the original sample surface, which allows the possibility of employing this method as a polishing technique[53], or to serially produce tiled membranes for large-area homoepitaxial CVD growth[54]. More relevantly for quantum technologies, the above-mentioned membranes provide the ideal substrate that guarantees vertical confinement of light in the subsequent fabrication of waveguides[55,56], photonic crystals[57,58] or ring resonator[59] structures, either with FIB techniques as described below or by electron beam lithography followed by reactive ion etching. Alternatively, the membranes can be employed without further microfabrication steps to take advantage of the either native or implanted color centers that they incorporate[60,61].

Since the crystallographic quality of the membrane is not always suitable for devices integrating individual color centers characterized by high coherence time, the lift-off technique has been combined with a subsequent CVD overgrowth phase, to create high-quality free-standing layers from which the original growth substrate could be removed via plasma etching. These CVD-grown membranes can be subsequently patterned and/or ion doped to produce optical/photonic cavities incorporating highly coherent color centers[12,62] (**Figure 3(d)**).

Taking advantage of the high dielectric strength of diamond, it is possible to apply intense electrical fields across lift-off-fabricated membranes with metal contacts and/or underlying conductive substrates, for the electrical stimulation and control of color centers[63].

Scanning keV FIB nanofabrication is finding an ever-increasing range of application for integrated photonic structures, thanks to significant progress in ion optics and innovative ion beam sources[64]. FIB techniques were employed to fabricate Solid Immersion Lenses (SIL) aligned with individual NVs in bulk diamond (**Figure 3(a)**), demonstrating an order-of-magnitude increase in light collection efficiency and allowing the single photon detection at ~Mcps rates[65–69]. FIB microfabrication also allowed the creation of air-clad[55] or ridge-type[70] waveguiding structures into free-standing diamond layers produced by means of the lift-off method, as shown in **Figure 3(b)**.

The high spatial resolution and fast prototyping of FIB fabrication were exploited to fabricate photonic crystals with cavity modes tuned to the emission of individual color centers[57,58,71–73]. To achieve vertical confinement, an undercutting technique is required, which can be achieved by employing angled FIB etching on bulk samples[71,72], or in free-standing membranes fabricated either by the lift-off method[57,58] or heteroepitaxial growth on Si substrates[73] (**Figure 3(c)**).

Another strategy to fabricate waveguides in diamond is to locally modify the refractive index through ion implantation. The irradiation process introduces a non-uniform vacancy density profile in the diamond substrate, with an end-of range peak where the crystal amorphization is most significant. The influence of the implanted ions on the local optical properties is in most cases negligible, especially in the case of light ions. The main challenge is to induce a sufficiently high local refractive index variation at end-of range through crystal amorphization, with only a limited increase in optical losses.

The first studies on the optical effects of ion implantation in crystals were performed by Townsend[74], who expressed the relative refractive index variation $\Delta n/n$ as a function of the relative variation of atomic density $\Delta N/N$ and polarizability $\Delta \alpha/\alpha$ as:

$$\frac{\Delta n}{n} = \frac{1}{6n^2}(n^2 - 1)(n^2 + 2)\left[\frac{\Delta N}{N} + \frac{\Delta \alpha}{\alpha} + f\right] \qquad (1)$$

where $f$ is a characteristic structure factor of the host material that is absent in cubic crystals[75]. Since ion implantation generally leads to a decrease in atomic density due to crystal amorphization and volume expansion, in most crystals it results in a decrease in $n$. However, in diamond, the role of polarizability is not negligible, due to the change in bonding character from $sp^3$ to $sp^2$.

Initial studies found an increase of refractive index $n$ in the visible spectrum for C ion implantations in keV and MeV ranges at low fluences, but a decrease of $n$ for higher fluences (above $10^{16}$/cm$^2$)[76,77]. More recently, several systematic studies highlighted a direct proportionality between the increase in the real and imaginary parts of $n$ and the induced vacancy density $\rho_V$, for proton implantations at MeV energies[78–80]. This holds true for other ion species and energies in the keV range, in the case of relatively small vacancy densities (~$10^{21}$/cm$^3$), for which the increase of the real part of $n$ at $\lambda = 637$ nm is on the order of 1% [81]. For larger vacancy density values, a saturation is found in the increase, and the data can be interpolated by an empirical exponential-like curve: $n = n_0 + n_\infty[1 - \exp(-\rho_V/b)]$, where $n_0$ is the refractive index of the undamaged diamond, $n_\infty$ the saturation value, and *b* a fitting constant. This has been verified using spectroscopic ellipsometry in a large spectral range (250-1750 nm)[81].

Further studies have reported partially contradictory results: Gregory *et al.* initially confirmed an increase in *n* in 1 MeV He-implanted diamond, using coherent acoustic phonon spectroscopy[82], but later amended their conclusions[83]. Draganski *et al.* observed a non-monotonic variation of *n* in 30 keV Ga-implanted diamond at different wavelengths, although results were affected by non-negligible uncertainties and the measured optical properties were likely influenced by the implanted ions themselves[84,85]. Despite these partial inconsistencies, the refractive index variation in ion-implanted diamond can be satisfactorily considered linearly proportional to vacancy density in the case of light species and low fluences, when optical properties are solely a function of crystal damage.

A summary of the main literature results is given in **Figure 4(a)**. All the studies highlight an increase in the extinction coefficient in implanted layers compared to pristine diamond over a large spectral range, so that low fluence implantations are in any case necessary to fabricate low-loss waveguides. The first buried waveguide based on refractive index modification in diamond by direct proton writing shown in **Figure 4(b),** was presented by Lagomarsino *et al.* [86]. Numerical simulations predicted the existence of guided modes (**Figure 4(c)**), which were then observed experimentally[86]. This study was followed up recently by Jin and coworkers, who used the same technique with a focused 2 MeV proton beam to demonstrate the fabrication of single-mode waveguides with a total insertion loss of 12 dB[87]. For these waveguides to be competitive with those created through other techniques, these losses need to be further reduced, possibly with thermal annealing treatments.

## 3.2 Femtosecond laser fabrication of photonic components

As with ion implantation, femtosecond laser writing leads to a reduced refractive index in crystalline materials, however this can be circumvented by inscribing two closely spaced lines to achieve waveguiding via the stress-optic effect. The first laser written waveguide in the bulk of diamond was achieved by a high repetition rate femtosecond laser amplifier[30]. The Yb:KGW system with 230-fs pulse duration, 515-nm wavelength was used for the fabrication. Polished 5 mm × 5 mm × 0.5 mm synthetic single-crystal optical grade diamond was employed.

Initial trials were focused on the creation of a single and continuous modification line in the bulk of diamond. Uniform and repeatable modifications in the bulk of diamond with 1.25 NA objective were found to be a pulse energy of 100 nJ, a scan speed of 0.5 mm/s, producing similar morphology at repetition rates of 5, 25 and 500 kHz. **Figure 5(a)** shows a cross sectional image of the laser-induced modification at 500 kHz repetition rate, revealing dimensions of 5 μm transversely and 20 μm vertically. As shown in **Figure 5(b)**, micro-Raman spectroscopy was performed on the laser written tracks to elucidate the structural modification [84,85]. The appearance of the G peak at 1575 cm$^{-1}$ and the D peak at 1360 cm$^{-1}$ indicate a transformation of sp$^3$ bonding into sp$^2$. The widths of the G peak being greater than 100 cm$^{-1}$ and the ratio of the intensities, $I$(D)/$I$(G) being close to 1, suggest that the sp$^2$ clusters mainly consist of amorphous carbon phase rather than graphite[88]. As the repetition rate was decreased, the G peak became sharper, indicating increased graphitization.

High resolution transmission electron microscopy (TEM) and electron energy loss spectroscopy (EELS) performed by the Salter group on femtosecond laser written tracks showed a non-uniform structural modification consisting of 4% of sp$^2$ bonded carbon[89]. The tracks were written using a femtosecond Ti:Sapphire laser at 1 kHz repetition rate. Despite the low amount of $sp^2$ carbon, the laser written wires showed good conductivity[90]. Further studies are needed to understand the role of the various laser parameters on the structure of laser written buried modifications in diamond. Having control on the amount of graphite within laser-written tracks is crucial not only for photonics where highly absorbing graphite should be avoided, but also microelectronics applications, where conversely, a high concentration of graphite is desirable.

*3.2.1 Femtosecond laser writing of bulk optical waveguides from the visible to mid infrared*

Motivated by the micro-Raman spectroscopy analysis revealing a small amount of graphite within laser induced tracks, particularly at the highest 500 kHz repetition rate, the type II writing method was exploited to demonstrate the first laser formed waveguide in the bulk of diamond[30]. Two lines written with 100 nJ pulse energy, 0.5 mm/s scan speed, 500 kHz repetition rate, 1.25 NA, 515 nm wavelength, and spaced by 13 µm enabled waveguiding in the visible, but only for TM polarization.

The waveguides were formed 50 µm below the surface and exhibited an insertion loss of 6 dB for a 5 mm long waveguide, with a mode field diameter (MFD) of 10 µm at 635-nm wavelength, as shown in **Figure 6(a)**. To gain further insight into the TM-only guiding, polarized micro Raman spectroscopy was applied on the type II waveguide from **Figure 6(a)**. By correlating the polarized Raman signal to the stress matrix around the type II waveguide, and considering the piezo-optic tensor of diamond, the refractive index profile map was obtained, as shown in **Figure 6(b)** for TM polarization. A positive refractive index change $\Delta n = 3 \times 10^{-3}$ at the waveguide center was found[91]. Conversely, for TE polarization, the refractive index decreases in the central guiding region, explaining the TM-only guiding of the waveguides.

The high refractive index of diamond and the large numerical aperture objective used for buried waveguide writing leads to strong spherical aberration at the focus, leading to a distortion of the intensity distribution, resulting in vertically elongated modifications. By applying a spatial light modulator (SLM), this aberration can be cancelled, leading to more symmetric modifications, providing more control over the waveguide cross section[92]. However, it was found without an SLM, and for depths of less than 30 μm, sidewalls having vertical dimension of 15 μm were produced, yielding single mode guidance in the visible[93]. For larger depths, where spherical aberration is more pronounced, the sidewalls were more elongated, leading to multiple guiding locations[30].

The wide transparency of diamond makes it a promising platform for near, mid and far IR applications. Waveguides with larger separation have been laser written for guiding of near - IR, telecom and mid-IR wavelengths. By increasing the spacing between the two lines, waveguides with progressively longer wavelengths were guided within a single mode. Modification lines with separations of 19 μm, 30 μm and 40 μm (**Figure 6(c)**) resulted in single mode guiding of 1.55 μm, 2.4 μm and 8.7 μm wavelength, respectively (**Figure 6(d)**). The mid IR wavelengths are promising for applications in molecular sensing, optical radar and astro-photonics[94].

*3.2.2 Femtosecond laser writing of advanced photonic components*

The demonstration of laser written waveguides in the bulk of diamond paves the way towards more complex photonic devices. Narrow-band reflectors enabling wavelength selective filtering are instrumental in quantum information[95], magnetometry[96] and Raman lasers[97]. Femtosecond laser writing was used to inscribe Bragg grating waveguides in diamond with 1.3 μm pitch for a 4th order Bragg reflection at 1550 nm wavelength[98]. A Ti:Sapphire femtosecond laser with 1 kHz repetition rate was applied to write the type II waveguide. An SLM provided a symmetric bulk modification and thus a multi-scan writing method was used to form the vertical sidewalls of the type II waveguide. The same laser system was used to form a periodic structure above the type II waveguide, and the resulting structure is shown in **Figure 7(a)**. The corresponding transmission and reflection spectra are shown in **Figure 7(b)**. Future work will seek to form periodic structures with reduced spacing to enable Bragg waveguides that operate at 532 nm and 637 nm wavelengths for use with quantum technologies based on NVs.

## 4. Deterministic placement of color centers

### 4.1 Ion beam implantation of color centers

Ion implantation enables the placement of individual color centers in diamond, allowing the introduction of specific ion species or isotopes in the diamond substrate, whose conversion into stable lattice defects occurs following high-temperature thermal treatment. Furthermore, the ion kinetic energy is correlated with its penetration range in the material, which allows fine control of the depth of the defects. Although ion implantation is a mature technique, the development of quantum devices relying on the deterministic fabrication of arrays of individual, indistinguishable single-photon sources with high spatial accuracy, requires further work.

*4.1.1 Fabrication*

**Color center creation:** The formation of single NVs can be achieved by irradiation with energetic protons, electrons or ions, inducing the formation of vacancies. These diffuse upon high temperature thermal annealing in the 600-1200°C range which promotes their combination with native nitrogen[101–104]. A recent set of studies indicate that temperatures higher than 1000°C result in better NV center spin coherence properties due to the reduction in the density of nearby irradiation-induced paramagnetic defects[103–105].

High purity quantum grade diamond with substitutional nitrogen concentration of ~1 ppb has a sufficiently low density of native NVs to enable controlled generation of single NVs. NV center ensembles on the other hand may be generated in diamond with a higher concentration of nitrogen, or through the implantation of nitrogen ions. In this case, ion implantation increases both the density of N atoms with respect to the native concentration and produces the vacancies needed for defect formation. The fabrication of high density of NV center ensembles requires high irradiation fluences, but care should be taken to avoid amorphization and subsequent graphitization of the diamond. For these reasons, the density of NV ensembles achievable upon room-temperature irradiations are bounded to values of ~$10^{18}$ cm$^{-3}$ (~10 ppm)[106]. The generation of alternative non NV color centers can also be achieved through the introduction of specified ion impurities in the diamond lattice.

**Nanoscale ion positioning:** The implantation of individual ions in diamond with accurate spatial resolution has been pursued by a variety of approaches. High positioning accuracy can only be achieved using ~sub-50 keV kinetic energies, to limit the ion straggling to few tens of nanometers[27,107]. Sub-100 nm positioning is highly challenging for general-purpose accelerators equipped with conventional electromagnetic focusing and scanning systems[100,108] and it is a demanding task even for keV FIBs, which demonstrated noteworthy results following the development of ad-hoc experimental setups[109–112]. FIB systems are sensitive to chromatic aberrations, resulting in a limited number of available ion sources and energies[27]. An achromatic alternative to achieve ion implantation with nanometer resolution consists of the collimation of a keV energy ion beam through properly defined apertures of 20-100 nm size. Apertures can be fabricated directly on the diamond surface (**Figure 8(a)**)[113,114], or they can consist of a separate collimating scanning system. In the latter case, different solutions have been explored, such as pierced AFM tips[115,116] (**Figure 8(b)**), a mica masks with nanochannels[117], or nano-stencil masks that are either custom scanned[118–120] or fixed[121–123] (**Figure 8(c)**).

**Single-ion delivery:** In conventional low-fluence implantation, the number of implanted ions is described by Poissonian statistics[102]. This approach has been used to produce single-photon emitters in diamond (**Figure 8(d)**) with irradiation fluences in the $10^8$–$10^{14}$ cm$^{-2}$ range, depending on the ion species and energy, thermal annealing parameters and substrate dopants concentration[124–126].

Two classes of strategies have been considered in recent years to move beyond this stochastic low fluence ion implantation to single ion delivery to ensure accurate alignment of color centers to photonic structures. The first strategy relies on pre-detection schemes, to ensure that an individual ion is present in the accelerating tube for delivery on the sample target. A practical implementation of such a system consists of the loading of individual ions in a trap, followed by their subsequent ejection in the accelerating tube[127–129]. This method enables in principle a fully deterministic implantation process and a highly monochromatic energy output as a consequence of the ion cooling in the trap. On the other hand, the number of available ion species compatible with such a system is still limited[130] and the implantation rate is constrained by the ion source loading and cooling processes. A complementary second strategy relies on the detection of ions on-the-fly during their path in the linear accelerating tube through the image charge principle[131]. However, the detection of an image charge associated with an individual ion, even in a highly-charged state, is a challenging task which has yet to be demonstrated.

In addition to both these strategies, post detection of ions embedded in the target material is needed to validate the deterministic placement process, as the accelerated ions might be scattered or screened by a nanometer-sized collimation system. The ion delivery detection can be used to provide an electrical trigger signal, feeding the activation of a beam blanker to prevent the implantation of additional ions. This task can be performed either through the detection of secondary electrons[132,133] produced by the impact of the impinging ions, or through the exploitation of the diamond substrate, equipped with suitable electrodes, as a solid-state single-ion detector[130,134]. In this latter case, the detection relies on the measurement of the induced charge signal formed as a consequence of the motion of the electron-hole cloud generated by the ion energy loss in the material[135]. Despite the large electron-hole pair energy creation of diamond (13.2 eV) in comparison to traditional detector materials, the approach has been successfully assessed for the room-temperature single-ion-detection at energies as low as 200 keV[134]. A further significant improvement in the sensitivity can be expected with the adoption of ion detection techniques at cryogenic temperatures and the development of high-sensitive, low-noise induced charge amplification chains.

**Formation yield**: The deterministic fabrication of diamond-based single-photon sources requires a high conversion efficiency of the implanted impurities into optically-active defects. Up until now, the only diamond-based emitters to meet this criterion are NV, silicon vacancy (SiV) and other group-IV color centers, as described below. However, the color center formation yield has been estimated on a statistical basis, relying on the measurement of the implantation fluence, due to the lack of convenient deterministic methods for the implantation of individual ions. The conversion efficiency of the NV center upon N implantation depends on the ion energy and the thermal annealing parameters, with values varying in the 1-25% range for 10-50 keV ions[105,136]. Similar yields were reported for the formation of the SiV centers upon keV Si implantation[112,137]. In both cases, the formation yield can be increased up to 25-50% by enhancing the surrounding vacancy density, either performing the ion implantation at MeV energies[102] or implementing an additional irradiation step to increase the vacancy density surrounding the N ions[99,138].

*4.1.2 Quantum emitters beyond the NV center*

The drawbacks of the NV center include a relatively long radiative lifetime, charge state blinking and broad spectral emission[139], which have stimulated research into alternative luminescent centers for solid state single-photon sources. Among them, the silicon-vacancy center (SiV, ZPL at 738 nm) has been extensively studied, despite its lower quantum efficiency and photo-stability at room temperature[140]. Its advantages are a tenfold shorter lifetime and a polarized light emission mainly originating from a narrow ZPL, which is lifetime-limited at cryogenic temperatures[141]. The SiV center can be regarded as the most promising optically-active defect in diamond for quantum information and quantum communication applications, due to an interesting degree of photon indistinguishability[141] and to the availability of schemes for the coherent control of its spin properties[142–145]. In recent years, additional emitters related to group-IV impurities have been explored, such as the germanium-related (GeV, ZPL at 602 nm), tin-related (SnV, ZPL at 620 nm) and lead-related (PbV, ZPL at 520 nm) color centers[13,14,124–126,146,147]. These defects are characterized by similar defect structure and opto-physical properties to those of the SiV center.

*4.1.3 Integrated photonic devices*

Ion implantation has been increasingly exploited in recent years for the fabrication of single-photon emitters embedded in integrated photonic structures. The preliminary development of such structures relied on the statistical implantation of individual color centers by means of unfocussed ion beams[148–150]. An alternative strategy consists of the fabrication of photonic structures such as solid-immersion lenses or photonic crystal cavity registered to a pre-characterized color center[67,71]. The recent availability of deterministic placement enabled the fabrication of individual NV centers[115,123,151] and SiV centers[112,152–154] (**Figure 8(e)**) at targeted positions of diamond photonic structures with high spatial accuracy. Nevertheless, the sub-optimal center formation yield requires the implantation of tens of ions to maximize the probability of obtaining an individual emitter, with a subsequent post-selection of the photonic devices required.

**4.2 Femtosecond laser written NVs**

*4.2.1 Femtosecond laser written single NVs and integrated waveguides*

Femtosecond laser writing has emerged in the last few years as an alternative method to ion implantation for the targeted creation of NV centers. It was shown in 2013 that NV center ensembles could be created at the surface of diamond through a femtosecond laser induced plasma, however this process caused ablative damage to the diamond surface[155]. The identification of GR1 neutral vacancy color centers in femtosecond laser written modification lines showed vacancies could be generated 10s of microns below the diamond surface[30]. To write single NVs in low nitrogen content diamond, the key was to find a parameter range where small numbers of vacancies are created, followed by high temperature annealing as noted in section 4.1.1. For the creation of integrated quantum optics devices incorporating single NV centers, it is also important to ensure that the annealing process does not cause any of the femtosecond laser written elements to degrade.

In 2016, Chen *et al.* demonstrated generation of arrays of single NVs at a depth of 50 μm in quantum grade diamond with substitutional nitrogen content of < 5 ppb, by writing single pulse femtosecond laser exposures, and annealing at 1000°C[156]. The use of a SLM allowed placement of the NV centers with an accuracy of better than 200 nm in the lateral direction, and 600 nm axially. This spatial resolution better than the optical diffraction limit is thanks to the nonlinear nature of the laser-material interaction[33]. The NV centers showed impressive optical and spin coherence properties, with photo luminescence excitation linewidths down to 12 MHz limited only by the Fourier transform of the excited state lifetime, and decoherence times exceeding 100 μs. **Figure 9(a)** shows an optical microscope image of a two-dimensional array of single femtosecond static exposure trials written in a concurrent study by the Eaton group, while **Figure 9(b)** shows an example confocal image of a single NV center created with 24 nJ pulse energy[93]. The PL spectrum (**Figure 9(c)**) displays the characteristic NV center ZPL at 637 nm, while the intensity autocorrelation measurement (**Figure 9(d)**) shows a $g^{(2)}(0) < 0.5$ demonstrating single photon emission. Note that emission from the neutral charged $NV^0$ at 575 nm is beyond the scale of the graph shown in **Figure 9(c)**. Previous work by the Oxford group has shown NV centers with no detectable $NV^0$ emission from laser writing and annealing with a similar setup (supplementary information[156]). Laser induced surface ablation on diamond has also been shown to create NV centers on the surface[157]. Although, the spin properties of NV is reduced for shallow NVs[158], these are promising for sensing applications.

More recently, the Salter group demonstrated three-dimensional arrays of single NV centers with coherence times exceeding 500 μs[159], and the generation of single NV centers with near unity yield through laser induced local annealing and online monitoring of the NV center fluorescence[160]. Such long coherence times suggest that minimal local damage associated with paramagnetic vacancy complexes is produced during the laser fabrication of NVs. In this work, a single seed pulse with energy 27 nJ to generate vacancies is followed by a 1 kHz train of lower energy (19 nJ) pulses, which stimulate local diffusion of the generated vacancies to generate NV centers. By monitoring the NV center fluorescence signal and continuing until an NV center is formed it is possible to removing the stochastic process of annealing in an external furnace. Remarkably the NV centers appear to retain long spin coherence (with measured coherence times up to 200 µs), even in a HPHT diamond sample with 1.8 ppm nitrogen content.

These techniques open the door to diamond chips with up to 5 million NV center qubits written solely through femtosecond laser writing. Such deep arrays have not been demonstrated through ion implantation which normally is restricted to near surface single color centers. However, for use in quantum computing, NV centers need to be coupled to each other through interaction with emitted single photons, motivating an attempt to combine single femtosecond laser written NV centers with previously described femtosecond laser written waveguides for a fully integrated single photon source device.

In order to achieve this, the Eaton group wrote waveguides in quantum grade diamond at 25 µm depth with a line of 5 single static exposures separated by 20 µm at the center of each waveguide. Following annealing at 1000°C, fluorescence confocal microscopy measurements confirmed that single NV centers had been created at several of the expected static exposure sites, and importantly the waveguides survived the annealing process[161]. Concentrating on NVs written with 28 nJ, it was shown that it is possible to excite the single NV centers by coupling a laser beam into the waveguide through a single mode fiber. Subsequently it was shown that NV center fluorescence could be collected through the waveguide by selectively exciting one of the single NV centers through the confocal microscope. Photoluminescence excitation studies of the NV centres at 6K demonstrated that although they did not have lifetime limited linewidths[162], they were comparable to the linewidths reported by the Salter group at the upper end of their laser writing energy range[156]. Previous measurements of as grown NV ensembles in waveguides showed that the NV centers retain their spin coherence[30]. This suggests that by further optimizing the writing energy, and annealing conditions or implementing the femtosecond laser localized annealing technique it should be possible to achieve these high quality NV centers within waveguides through femtosecond laser writing. This proof of principle waveguide quantum photonics device demonstrates the easy compatibility of femtosecond laser written devices with existing fiber technology, however the NV center collection efficiency was limited by the relatively large effective mode area of the 9.5 μm MFD waveguide. This can be increased by reducing the mode field diameter of the waveguides, however this should be achieved without compromising the loss rates of the waveguides. Another approach which utilizes larger MFD waveguides is to combine the waveguides with high density ensembles of NV centers for broadband NV center enabled sensing applications.

*4.2.2 High density writing of NVs for sensing*

The NV center can be used as a room temperature quantum sensing resource by measuring the shift of the ground state spin transition frequencies under the influence of magnetic and electric fields or changes in temperature through optically detected magnetic resonance measurements (ODMR)[96,163–167]. NV sensing using single NVs has been demonstrated for example using single NVs in nano-diamonds for nanoscale atomic force microscopy based magnetometry[167], or single centers near the surface of bulk diamond for electric field sensing of a single fundamental charge[166]. In these demonstrations, the atomically small size of a single isolated NV center is used to allow for nanoscale spatial resolution surpassing the resolution of the optical fluorescence microscopes used for measurement, while the required measurement sensitivity is achieved by averaging over many measurements, since the detected fluorescence rate is limited by the use of a single NV center.

For higher speed (or high sensitivity) sensing applications which do not require such nanoscale spatial resolution, single NV centers may be replaced with high density NV center ensembles[96]. Such high sensitivity broadband ensemble NV magnetic[3] and electric field sensing[168] has been demonstrated in bulk diamond which has been ion irradiated or has high as grown NV densities. The sensitivity for an ensemble of NV centers can be described by

$$\eta = \frac{1}{k} \frac{1}{C\sqrt{M\Gamma}} \frac{1}{T_2^*}, \qquad (2)$$

where $k$ represents the frequency shift of the NV center's spin transitions under magnetic ($k_m$ = 28 GHz T$^{-1}$) or electric fields ($k_e$ = 17 Hz cm V$^{-1}$)[169], $C$ is the ODMR contrast, $M$ is the number of NV centers in the ensemble measurement, $\Gamma$ is the detected count rate for each NV center, and $T_2^*$ is the inhomogeneous NV coherence time[170]. The measurement sensitivity increases by a factor of $\sqrt{M}$ as the number of NV centers in the ensemble increases, which can be achieved by either increasing volume of NV centers or the density of NV centers sampled. However, NV center density should not be increased to the point where the coherence time of the NV centers is reduced[96]. As described in section 4.1.1, ion or electron beam implantation followed by annealing have both previously been used to generate high density NV center ensembles for sensing applications, with NV densities ranging from $3\times10^{13}$ cm$^{-3}$ - $1\times10^{18}$ cm$^{-3}$ (0.2 ppb - 6 ppm) in CVD and HPHT grown diamond[168,171–173]. Clevenson *et al.* identified an optimal NV density (without a serious reduction in coherence time) of ~$2\times10^{16}$ cm$^{-3}$ (0.1 ppm)[3].

A previous study for the generation of femtosecond laser written ensembles in diamond with nitrogen content ~ 100 ppb showed it was possible to write NV ensembles with NV density of ~$5\times10^{13}$ cm$^{-3}$ (0.3 ppb), by writing single static exposures, followed by annealing[174]. Similar conversion efficiency should be possible to achieve NV ensembles with density of ~$5\times10^{16}$ cm$^{-3}$ (0.3 ppm) in HPHT diamond with nitrogen content of around 100 ppm.

To investigate the creation of an integrated waveguide device containing a high-density NV center ensemble, waveguides were written in HPHT diamond with a mode field diameter of 10 µm at a depth of 18 µm. 'Empty' waveguides were written alongside 'static exposure' waveguides which contained arrays of 9 single 100 nJ static exposures separated by 2 µm axially, and with a pitch of 1 µm along the waveguide to generate NV centers in the waveguide mode. Cross-section schematics of the 'empty' and 'static exposure' waveguides are shown in **Figure 10(a) and (b)**, respectively.

**Figure 10(c) and (d)** show PL confocal maps in transverse view (upper), and overhead view (lower) of the 'empty' and 'static exposure' waveguides. Intriguingly, although bright spots corresponding to static exposures are visible in the 'static exposure' waveguide, a broad bright fluorescence is observed in both waveguides. PL spectra taken from the 'empty' and 'static exposure' waveguides, and from a bright spot corresponding to static exposure in the waveguide after annealing are shown in **Figure 11(a)**. The spectra show the characteristic ZPL of the NV center at 637 nm, suggesting that vacancies created during the inscription of the modification lines have diffused into the center of the waveguide to form NV centers. The spectra taken from the static exposure is even brighter than the surrounding area and shows additional structure with lines at 660 nm, 680 nm, 697.6 nm, 720.7 nm, 741.1 nm, 760.7 nm. It also appears to show a much stronger contribution from a line at 575 nm corresponding to the neutral $NV^0$ ZPL.

In order to make an estimate of the NV density in the waveguides, power dependent PL saturation measurements of the 'empty' and 'static exposure' waveguides were performed and compared with the total signal collected from a single NV center in another sample measured using the same conditions (**Figure 11(b)**). Similarly, it is possible to make an estimate of the collection volume of the confocal microscope using confocal images of a single NV to approximate the confocal microscope's point spread function. Thus, we find that with a 0.8 NA objective and a confocal collection volume of ~1 µm³, we have a NV center density of ~1.1×10¹⁵ cm⁻³ (or 6 ppb) in the 'empty' waveguide, and up to ~1.4×10¹⁵ cm⁻³ (or 8 ppb) in the static exposure waveguide, although further characterization of the NVs in the waveguides and static exposures is required to clarify the coherence times of the generated color centers.

Using these estimated NV center densities, we can make a projection of the possible NV center electric field and magnetic field sensitivities we could expect from a femtosecond laser written waveguide sensing device with excitation and detection performed through coupling to the waveguide such as illustrated in **Figure 11(c)**. A 3 mm long waveguide with a 10 µm MFD, and NV density of $1.1 \times 10^{15}$ cm⁻³ can expect to have $M = 2.6 \times 10^8$ NV centers contributing to the sensing signal, with a photon collection efficiency of $\Gamma = 900$ Hz per NV and a contrast $C = 0.05$. This suggests a magnetic field sensitivity of 1.5 nT Hz$^{-1/2}$ or alternatively an electric field sensitivity of 2.4 V cm⁻¹ Hz$^{-1/2}$, assuming the NV centers have $T_2^* = 1$ µs, as has been demonstrated in similar density NV center ensembles[3,168]. Although this is above the state-of-the-art sensitivities achieved for magnetic (~290 pT Hz$^{-1/2}$ [3]), and electric field (~1.6 V cm⁻¹ Hz$^{-1/2}$ [168]) sensing, it should be possible to improve through optimizing the material and femtosecond writing parameters to further increase the achieved densities. It may also be possible to improve the sensitivity even further by integrating femtosecond laser written Bragg reflectors into the waveguides to increase the effective light-NV center interaction length through multiple passes of the excitation laser. Finally, the strain within the waveguides which enables waveguiding[91] opens interesting possibilities for improving electric field sensing in particular. Since strain is equivalent to an electric field, the initial strain which each NV center in the waveguide experiences should act as a bias electric field, allowing for the NV centers to be less affected by stray magnetic fields, and more sensitive to extremely weak electric fields.

## 5. Microfluidic channels in diamond

Microfluidic systems are devices characterized by channels with transverse dimension of typically ~100 µm for the manipulation of fluids with volumes below the microliter range. These systems allow miniaturization and integration of complex functions, suggesting the possibility to perform numerous experiments rapidly and in parallel while consuming small amounts of reagents[175–177]. Microfluidic devices can integrate analytical detection techniques[178,179], such as electrochemical and optical methods and represent a perfect tool for biological application both as a supporting system in conventional biosensing devices (i.e. perfusion)[180] or an active component for cell activity monitoring and biomolecule analysis[181–183].

Diamond is an attractive substrate for microfluidics since it offers high chemical resistance (for applications with highly corrosive reagents), high mechanical strength (for high-pressure pumping), transparency (for optical probing) and high thermal conductivity (for solution heating or thermal dissipation). Moreover, it offers integration of microfluidic channels with advanced quantum sensing schemes based on color centers hosted in the diamond. This cutting-edge approach has demonstrated field distribution mapping and magnetometry with DC sensitivity of 9 µT Hz$^{-1/2}$ in fluids[184–186]. An example device geometry proposed by Ziem *et al.* is shown in **Figure 12(a)**[187]. Alternatively, a hybrid PDMS – diamond biosensor for the detection of charged molecules through NV charge state monitoring is shown in **Figure 12(b)** [188]. Microfluidic structures in diamond have been demonstrated using standard lithographic techniques[189–191], epitaxial lateral overgrowth[192], laser ablation[193] and ion beam lithography[194,195] (**Figure 12(c)-(e)**). The latter two approaches will be described in detail below.

**5.1 Micro-channels written with Ion beam implantation**

MeV Ion Beam Lithography was employed for the fabrication of microfluidic structures with a monolithic approach, based on graphitization and selective etching. As previously described, ion implantation damage localized at the ion end of range (for MeV light ions a few microns below the surface) induces diamond conversion into an amorphous phase, which subsequently converts to nanocrystalline graphite after thermal treatment (>800°C)[28]. The final effect is the creation of a thin (100 nm – 250 nm) graphitic layer with thickness determined by the ion implantation damage profile. A uniform "box-like" damage density profile can be created by adopting a multiple-energy implantation approach, which enables the extension of the implanted layer along the depth direction by tuning the implantation energies and fluences, allowing the creation of channels with optimal thicknesses for applications in microfluidics.

Two different strategies can be used for implantation: a microbeam or a collimated ion beam. The former approach employs ion beams with a spot dimension of few µm raster scanned on a defined pattern for direct writing of the channels. The latter approach uses a broad beam collimated by employing a metal mask, which defines the geometry of the channel with micrometric spatial resolution. This guarantees the parallelization of the creation of the structure and it represents an optimal solution to perform multiple-energy irradiation of the desired regions since it avoids beam refocusing and sample misalignment. On the other hand, the uniformity of the beam becomes a crucial parameter for the reproducibility of the lithographic process.

Since the obtained graphitic structures are fully embedded into the diamond matrix, holes need to be created in order to reach them. FIB milling or high-power laser ablation are fabrication techniques capable of removing diamond in a controlled manner. FIB creates apertures with higher resolution and definition of the desired geometry, but it is a time-consuming process.

Femtosecond laser milling is more practical for creating access holes with a larger diameter that can be quickly and efficiently made and interfaced with microfluidic pumps[194]. Finally, selective removal of the graphite with respect to the surrounding diamond matrix is obtained by oxidation in order to create hollow microfluidic channels. In the literature, three different strategies have been adopted: high-temperature dry ozone or oxygen etching; hot acid etching; and electrochemical etching. High temperature dry etching of the graphite in ozone or oxygen is performed by heating the substrate to a temperature where the reactive gases oxidize carbon in the graphite phase producing carbon oxide or carbon dioxide. Selective etching of graphite is achieved in a temperature range of 380 – 550°C, avoiding the diamond removal which occurs at higher temperatures. Both the etchant ($O_2$ or ozone) and the etch products (CO and $CO_2$) are gases, resulting in a very clean process, and the diffusion of gaseous phase material in the thin graphitic region is more rapid than for a wet process. Wet strong heated acid etching has also been demonstrated, typically using a $H_2SO_4/NaNO_3$ mix or a 1:1:1 solution of $HNO_3/H_2SO_4/HClO_4$, which can penetrate the channels by capillarity[49]. Electrochemical etching is achieved by the immersion of the sample into an electrically conductive solution (distilled water and $H_3BO_3$ with a final concentration of 1-5 × $10^3$ mol $L^{-1}$) while a DC voltage of 150-200 V is applied through a pair of platinum electrodes placed in proximity to the sample. The distance between electrodes and graphitic structures influence the intensity of the applied bias. The graphite structures are polarized by the electric field creating a virtual anode and cathode on the sample. Anodic oxidation reactions occurring at the edge of the implanted layer result in the selective etching of the graphite with respect to the diamond matrix[196] (**Figure 12(d)-(e)**).

MeV Ion beam lithography represents a useful tool to realize lab-on-a-chip devices since it guarantees the creation of channels with sub-micrometric resolution in all three dimensions while preserving the optical properties. However, further work is required to overcome the dramatic insurgence of cracks into the cap layer due to the residual stresses caused by the process.

**5.2 Surface patterning using Bessel beams**

*5.2.1   Surface microfluidic channels*

Standard laser micromachining techniques making use of focused Gaussian beams leads to a high degree spatial energy confinement with an aspect-ratio influenced by the dimensions of the focal volume. Laser microfabrication with this approach imposes a constraint on the speed and quality of the microfabrication in some applications where larger dimensions of the machined structures are required. For instance, in microfluidics, extended channel depth allows for increased throughput at the optimized flow velocity[197]. Nonconventional beam shapes are increasingly under investigation to better tailor the laser micropatterned features to meet the requirements of a given material configuration or application[198].

In particular, it was recently shown that laser microfabrication making use of ultrashort Bessel beams (BB) could be applied to achieve deep surface ablation of diamond in a single pass[193,199]. The results are based on the fact that the intensity distribution of a Bessel beam has a central core surrounded by rings which constitute the beam energy reservoir for a non-diffracting propagation (**Figure 13**). In particular, in the stationary nonlinear regime, a finite energy Bessel beam leaves in its wake a uniform elongated plasma track, generated by the main Bessel lobe, that is the main support for the nonlinear absorption of laser energy.

The Bessel beam has thus been used in transverse writing configuration in such a way to generate an ablation pattern on the diamond substrate surface with micron-sized high aspect-ratio trenches opportunely tailored as a function of the beam parameters and writing speed[193]. The central part of the written track whose depth below the surface (up to 20-30 μm) depends on the pulse energy, with a homogeneous and smooth microchannel created by the BB core during the nonlinear absorption process, as shown in the scanning electron microscopy (SEM) images in **Figure 14**. Atomic Force Microscopy further highlighted the presence of nanogrooves inside the ablated microstructures. The micro and nanofeatures of the generated surface microchannels can be tailored as a function of the laser beam parameters, making this fabrication technique interesting for the fast writing of channel-like microstructures potentially useful for microfluidics applications, with sensing functionality provided by NV centers.

*5.2.2 Generation of pillar-like microstructures*

Writing nanometer-sized substructures in a controlled way inside the trenches of a microfluidic chip can give added value especially for biosensing applications. Within this context, the possibility to tailor the geometry and features of the surface microchannels that can be generated on diamond as a function of the BB cone angle used, the pulse duration and the pulse energy, was further studied in Kumar *et al*.[199]. Moreover, an increase of the area to be functionalized is often performed by fabricating micropillars on the material surface[193,199], these being typically obtained by chemical etching lithography or ion beam lithography, and thus by means of lengthy processes. Closely spaced 3D microstructures with tailorable features and different shapes (pillar-like or tip-shape) were realized by laser machining the diamond surface in transverse configuration by orthogonally crossing the writing trajectories obtained with BB, and with varied laser writing parameters, with the advantage of a faster processing. Resulting SEM images are shown in **Figure 15** for illustration[199].

The morphology of these microstructures strongly depends on that of the ablated surface trenches. Pillars which present smooth and regular surfaces, with vertical walls can be created. By increasing the cone angle, it is also possible to obtain more tip-like microstructures with a modulated surface pattern as in **Figure 15(a) and (b)**, pillar-like microstructures where nanofeatures entirely characterize the surfaces as in **Figure 15(c)** and finally cubic microstructures with a flat surface (**Figure 15(d)**). In general, these results validate the use of this laser machining technique for an ad-hoc micro-structuring process of biocompatible diamond surfaces in all cases where well-defined microstructures with controlled roughness are needed, for instance for microfluidics or sensing, and thus for all cases where an increase of the surface area to be functionalized is required.

## 6. Conclusions and Future Trends

Overall, energetic (keV – MeV) ions beams and femtosecond laser writing have both proved to be powerful and versatile tools to modify the structural, optical and electrical properties of diamond for applications in quantum-optical devices. The key advantages and limitations of ion beam techniques can be regarded as complementary those of femtosecond laser writing. Firstly, ion beam lithography in diamond can rely on an extensive body of previous work, dating back to the 1970s, with pioneering demonstrations of ion-induced graphitization[200] building on the consolidated use of ion beam irradiation in the semiconductor industry. This translates to the availability of well-consolidated practices and protocols, in comparison to the more recently developed laser fabrication, which is rapidly developing new techniques and modalities. Ion beam fabrication can offer versatility through the employment of different ion species, that grants a further degree of freedom in the engineering of the structural modification of the material (penetration depth, damage density, etc.) with respect to the mere variation of the primary radiation energy, and most importantly allows the doping of the target crystal with foreign impurities with a high level of control. This advantage, however, must be weighed against the accessibility and cost of instruments: although remarkable efforts have been made to develop compact and user-friendly apparatuses, it is fair to say that a multi-species and variable-energy ion implanter is a relatively expensive and technically demanding experimental setup, with complex operating and maintenance procedures with respect to laser setups, particularly if designed to operate with high spatial resolution.

On the subject of spatial resolution, and more generally on the geometrical features of the structures that can be fabricated with the two approaches, again elements of complementarity emerge. On the one hand, with ion beam irradiation it is possible to fabricate graphitic structures which are generally more regular in their shapes, particularly in terms of roughness: typical graphitic structures produced with this method are smooth at the nanometer scale[201]. This feature is due to the finely controlled radiation-induced damage density allowed by the possibility of easily setting the fluence delivery rates across several orders of magnitude. On the other hand, ion beams are significantly less flexible than laser beams in terms of the definition of the geometries of the structures that can be fabricated, particularly in terms of their depths within the bulk of substrate: although determined by the ion energy and species, such depth can vary over a rather limited spatial range, at the cost of modifying the above-mentioned parameters, i.e. a task which is not as straightforward in its technical implementation as changing the focal depth in laser irradiation systems. Similar to the penetration depth, the spatial resolution in ion lithography is also affected by fundamental limitations that are inherently based on the nature of the radiation-material interaction, i.e. on the non-negligible lateral and longitudinal straggling processes. Straggling phenomena can be minimized with the use of low-energy ions, unfortunately leading to the creation of shallower color centers, which are generally characterized by lower creation efficiencies. Although ingeniously addressed with several technological solutions in recent studies, this fundamental limitation currently represents a challenging barrier to overcome, in order to reach the ultimate spatial resolution necessary for the implementation of realistic quantum devices. Femtosecond laser writing of color centers, on the other hand, is currently limited by the spatial resolution both laterally and vertically.

A key feature in which ion beam fabrication faces substantial limitations is the creation efficiency of the color centers. Despite considerable research efforts in the past decade, the ratio of quantum-optically functional color centers per implanted ion is still largely unsuitable for the scalable fabrication of multiple-qubit devices. This is due to a limited understanding and thus control of the complex dynamics that govern the formation of defects during ion irradiation (e.g.: defect-defect interaction, self-annealing processes), as well as their recombination during the subsequent thermal annealing (e.g. defect migration, Frenkel pair recombination, formation of undesired defect aggregates). This compares on the other hand to the recent demonstration of an almost 100% creation efficiency for femtosecond laser written NV centers through fluorescence monitoring, and the creation of NV centers with excellent optical and spin coherence qualities.

In perspective, significant advances can be foreseen in both fields of ion- and laser-irradiation for the fabrication of scalable quantum devices in diamond. As far as the former field is concerned, new and compact ion sources are under development to extend the availability and user-friendliness of different ions species over increasing energy ranges, and innovative solutions to improve both spatial resolution, deterministic single-ion implantation and color center creation efficiency are under active investigation. Furthermore, we foresee that in the near future new appealing possibilities will be enabled by the integration of the two techniques into hybrid fabrication protocols, possibly by means of integrated experimental setups in which (for example) laser-induced modification of the material can initiate, progress or extend from graphitic seeds formed by ion irradiation[202], thus taking full advantage of the features of both methodologies.

**Acknowledgements**

S.M.E., V.B. and J.P.H. contributed equally to the work. The authors thank Patrick Salter and Angelo Bifone for fruitful scientific discussions. The researchers at IFN are grateful for support from the SIR MIUR grant "DIAMANTE" and H2020 Marie Curie ITN project "PHOTOTRAIN". IFN thanks Prof. Guglielmo Lanzani and Dr. Luigino Criante for the use of the FemtoFab facility at CNST-IIT Milano. R.O. acknowledges financial support from the European Research Council (ERC) under the European Union's Horizon 2020 research and innovation programme (project CAPABLE – 742745 – www.capable-erc.eu). The researchers at UniTo and INFN gratefully acknowledge the support of the following projects: Project "Piemonte Quantum Enabling Technologies" (PiQuET), funded by the Piemonte Region within the "Infra-P" scheme (POR-FESR 2014-2020 program of the European Union)"; "Finanziamento ex-post di progetti di ricerca di Ateneo" funded by the University of Torino with the support of the "Fondazione di San Paolo"; "Departments of Excellence" (L. 232/2016), funded by MIUR; "DIESIS" and "DIACELL" projects funded by INFN; coordinated research project "F11020" of the International Atomic Energy Agency (IAEA); projects 17FUN06 "SIQUST" and 17FUN01 "BeCOMe" funded by the EMPIR Programme cofinanced by the Participating States and from the European Union Horizon 2020 Research and Innovation Programme.

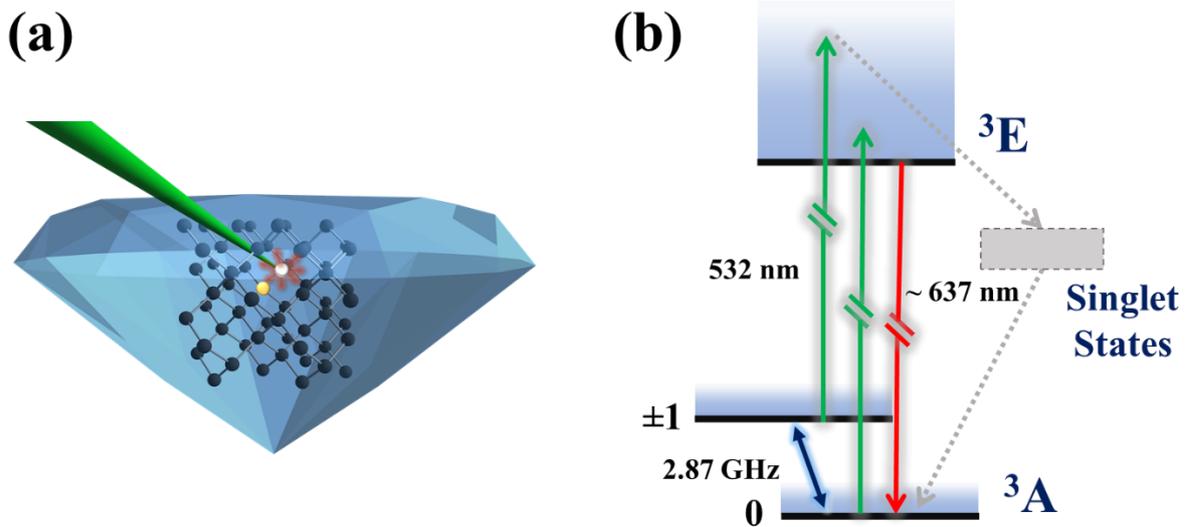

**Figure 1.** (a) Schematic showing the crystal structure of diamond with NV center (Carbon atoms represented in black, nitrogen atom shown in yellow and the vacancy shown in white), excited with a green laser beam and the vacancy is shown to fluoresce in red. (b) Simplified electronic energy level diagram of the ground state of NV center showing the triplet ground ($^3A$) and excited states ($^3E$), and the metastable singlet manifold. Optical transitions are indicated by arrows.

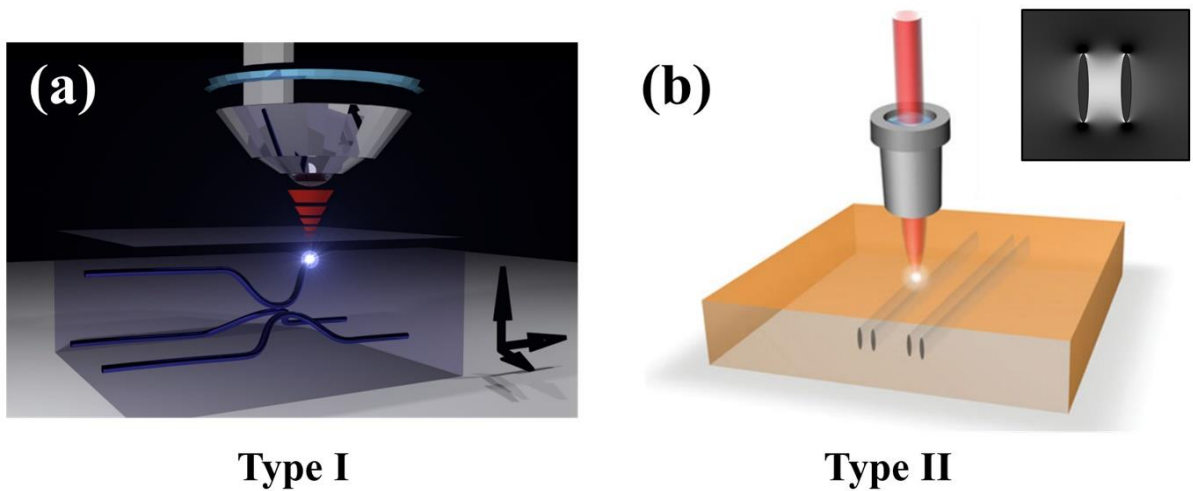

**Figure 2.** (a) Type I waveguide writing: focused fs laser pulses are nonlinearly absorbed below the surface of a transparent material resulting in a permanent refractive index increase[203]. (b) Type II: two lines of reduced refractive index are laser written close together, producing a stressed central region capable of guiding light. In both modalities, motion stages can be used to translate the sample with respect to the beam to produce 3D photonics[36].

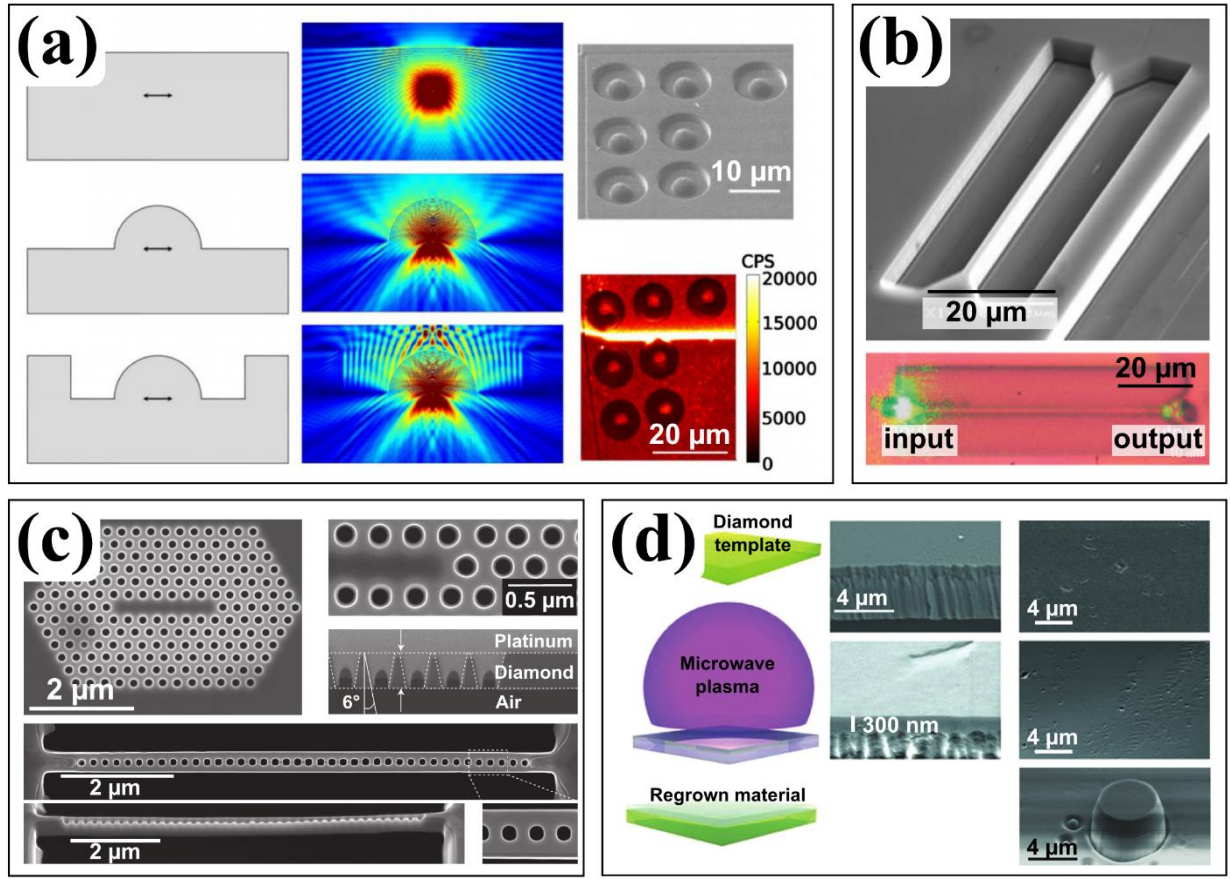

**Figure 3.** Photonic structures fabricated in diamond with ion-beam-based techniques: (a) Solid Immersion Lens fabricated with FIB in single-crystal diamond to improve light collection efficiency from photoluminescent emitters localized in the bulk[69]; (b) Suspended waveguiding structures fabricated with a FIB-assisted lift-off technique[55]; (c) Photonic structures fabricated with FIB in free-standing membranes removed from Si substrates on which they were heteroepitaxially grown[73]; (d) Free standing membranes and relevant integrated microstructures obtained by the lift-off of homoepitaxially CVD grown films[12].

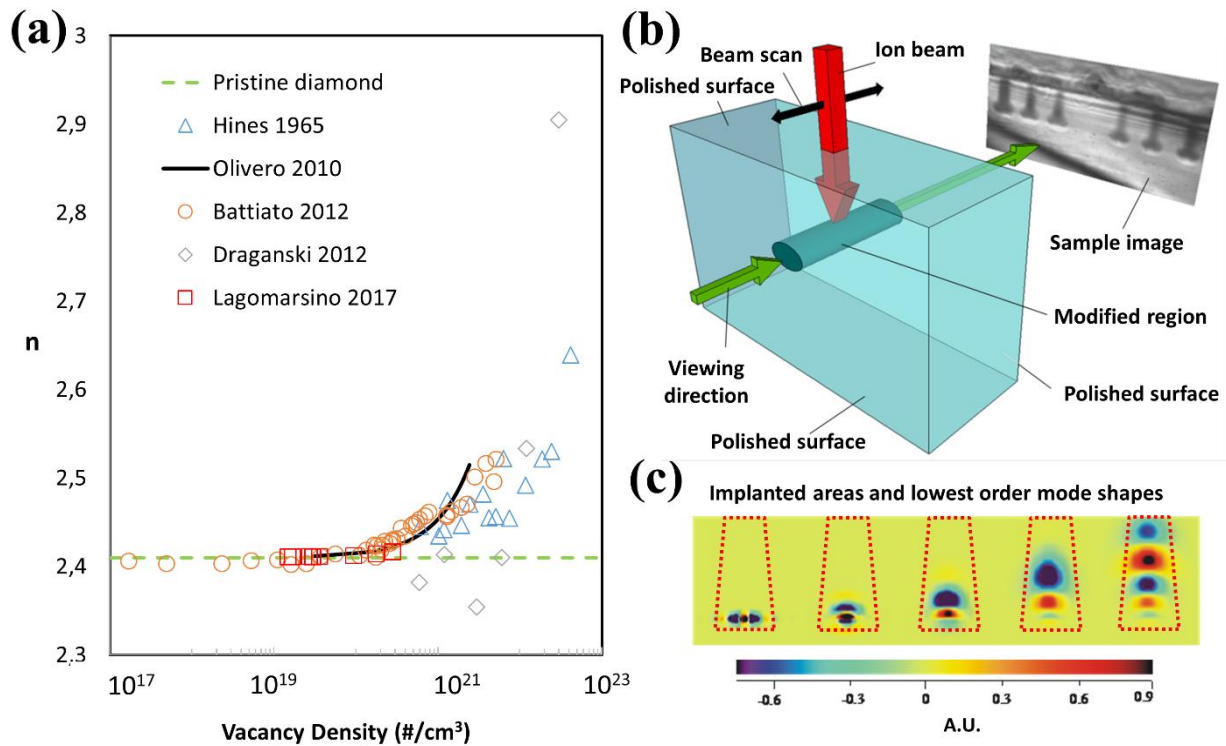

**Figure 4.** (a) Comparison between available results in the literature for the variation of refractive index in ion implanted diamond as a function of vacancy density at $\lambda = 638$ nm[77,79–81,84]; (b) Schematic of the fabrication and measurement geometry for the proton beam-written waveguides in Lagomarsino et al.[86]; (c) Simulated amplitude maps of the first 5 propagating modes in the waveguide[86].

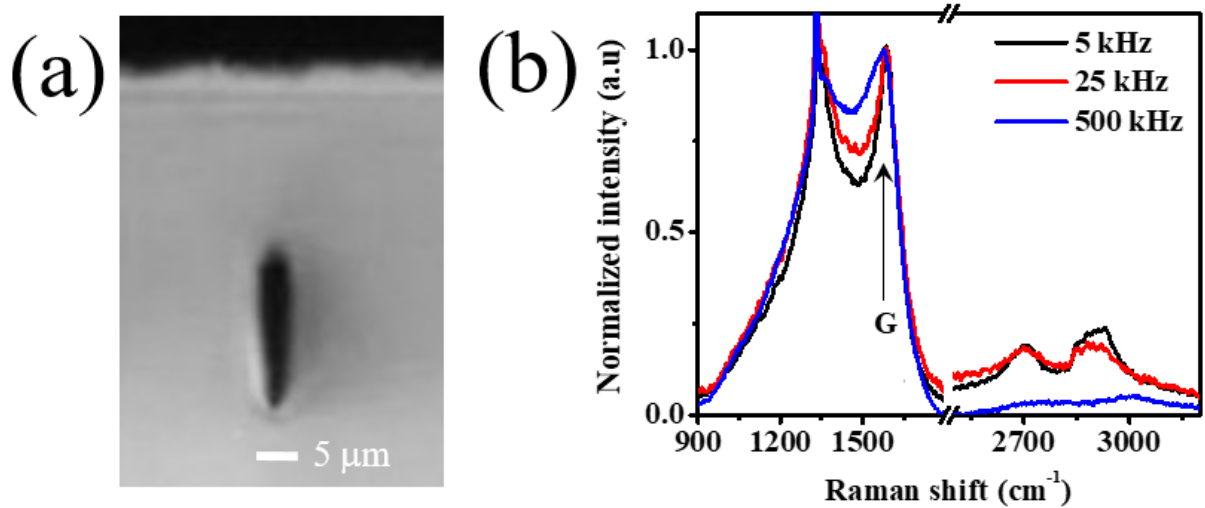

**Figure 5.** (a) Cross sectional transmission optical microscope image of a single laser written track written at 50 µm depth in diamond at 500 kHz repetition rate[30]. (b) Micro-Raman spectra captured within the laser written track at repetition rates of 5, 25 and 500 kHz[30].

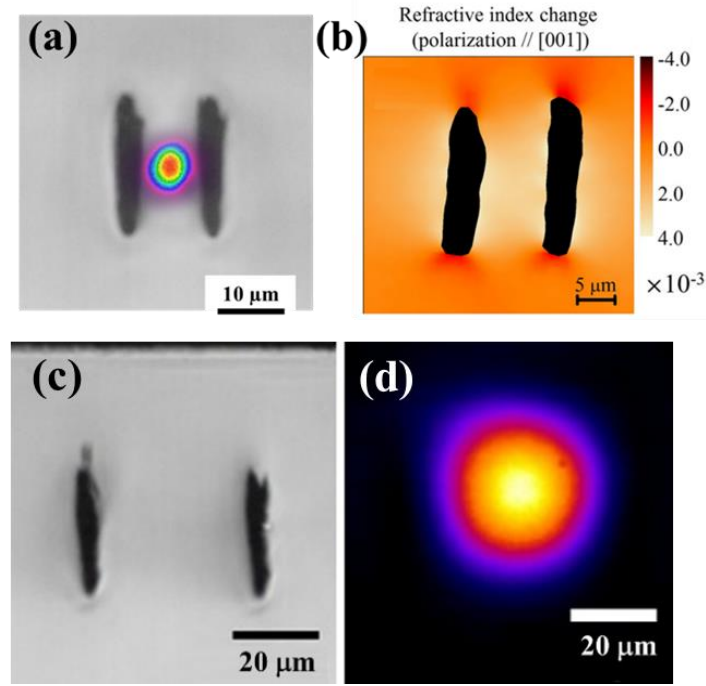

**Figure 6.** (a) Cross sectional transmission optical microscope image of type II waveguide with a separation of 13 μm and guided mode at 635-nm wavelength shown[30]. (b) A map of refractive index profile within the waveguide, obtained from polarized micro-Raman analysis ([001] crystallographic direction is taken along the vertical axis)[91]. (c) Transverse optical microscope image of type II waveguide with a separation of 40 μm and (d) the optical guided mode at 8.7 μm[94].

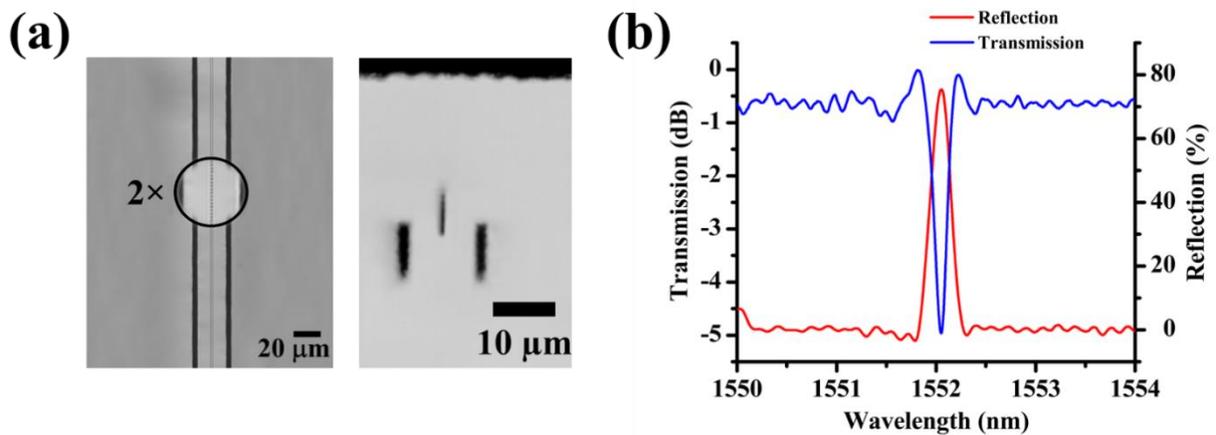

**Figure 7.** (a) Overhead (left) and transverse (right) transmission optical microscopy image of Bragg grating waveguide inscribed in diamond[98]. (b) Transmission (blue) and reflection (red) spectra of the Bragg grating waveguide[98].

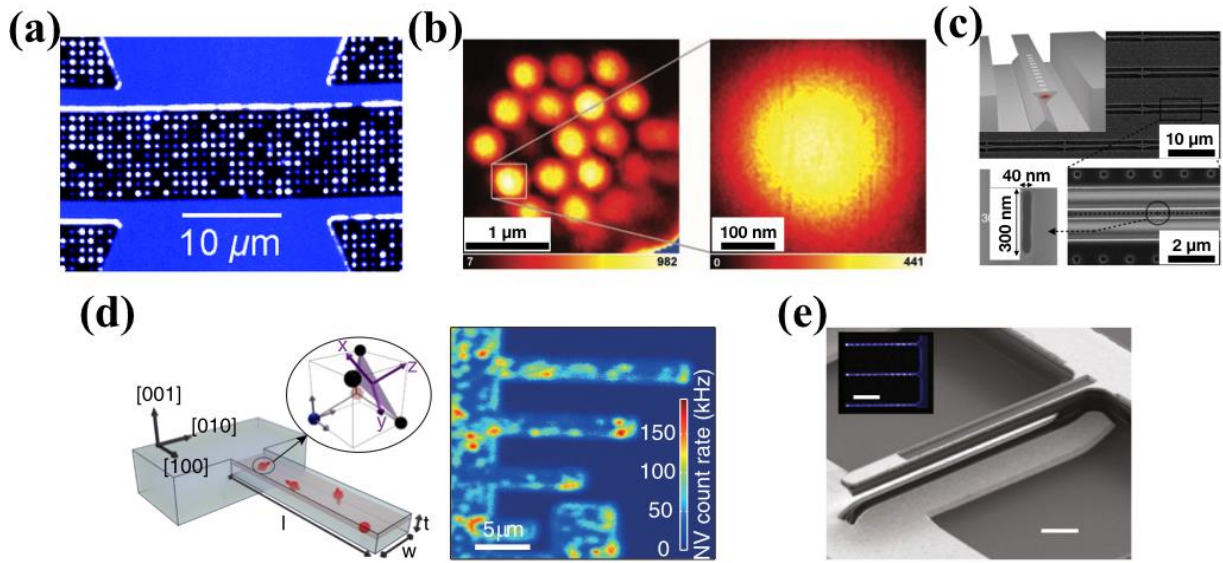

**Figure 8.** Integration of diamond-based single-photon emitters in diamond structures. (a) Large-scale fabrication of an NV center array by N implantation through a patterned resist mask[114]. (b) High-resolution fabrication of NV centers' by means of implantation through a pierced scanning AFM collimator[116]. (c) 1-dimensional photonic cavity equipped with an embedded NV center's array fabricated through targeted FIB N implantation[123]. (d) NV centers embedded in an opto-mechanical resonator through a broad N beam implantation at low fluence[150]. (e) Diamond electro-mechanical system consisting of a cantilever with SiV centers fabricated by FIB ion implantation[153].

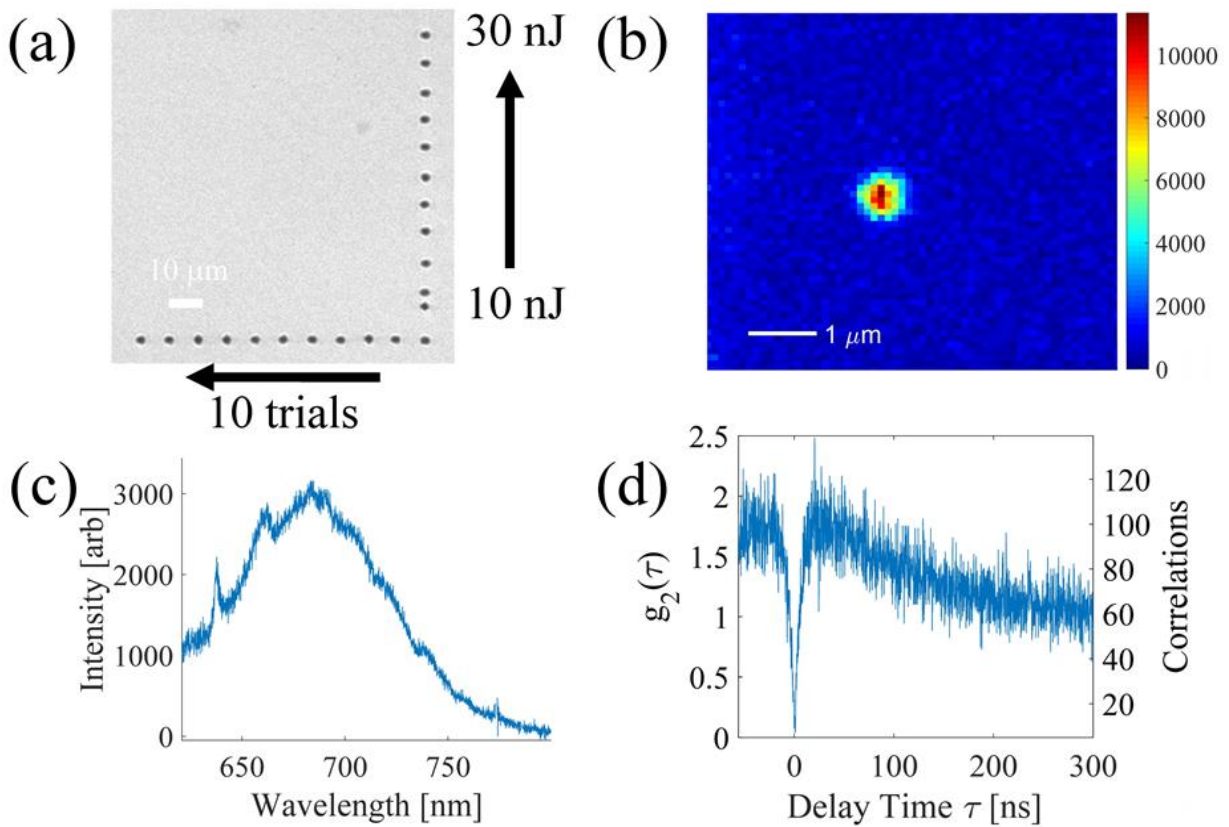

**Figure 9.** (a) Overhead optical microscope image of 2D array of single femtosecond laser exposures. Markers visible in both the optical microscope and the fluorescence microscope to enable orientation of the sample have been written at the edge of the array region with higher

fluence. (b) Overhead confocal PL intensity map and (c) spectrum from a single NV produced by a single 24 nJ ultrashort laser pulse followed by annealing. (d) Intensity autocorrelation measurement of photons from the single NV revealing single photon emission[93].

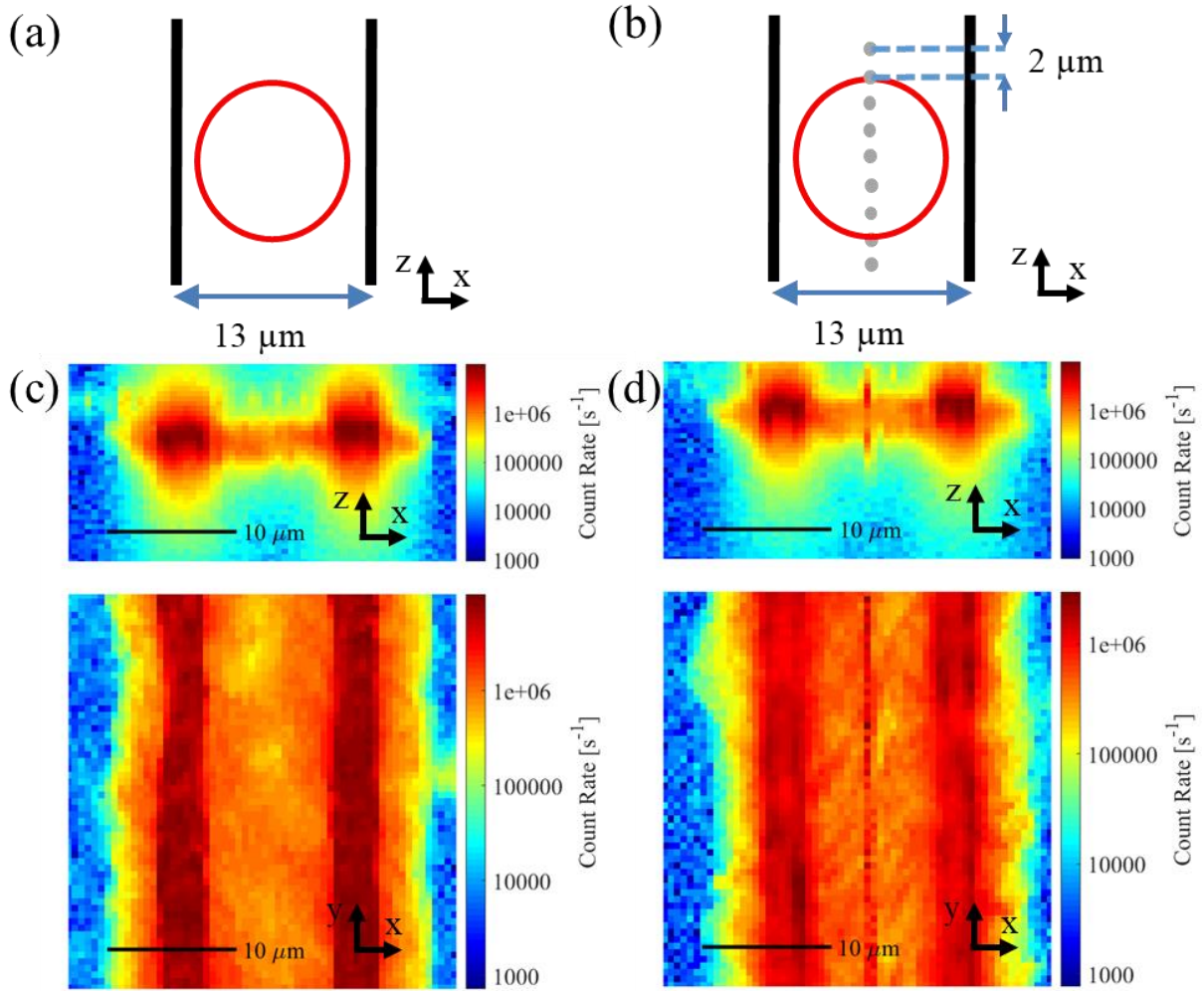

**Figure 10.** Schematic cross-section of femtosecond laser written (a) 'empty' and (b) 'static exposure' waveguides, with modification lines in black separated by 13 µm and the expected 10 µm mode field diameter of the waveguide mode outlined in red, with integrated static exposures, separated by 2 µm axially marked in gray. Photoluminescence confocal cross-sections (above) and maps (below) of (c) 'empty' and (d) 'static exposure' waveguide and modification lines written in HPHT sample after annealing. PL maps were acquired from the top of the sample using a homebuilt confocal microscope, causing an apparent axial squeezing of the cross-section due to refraction.

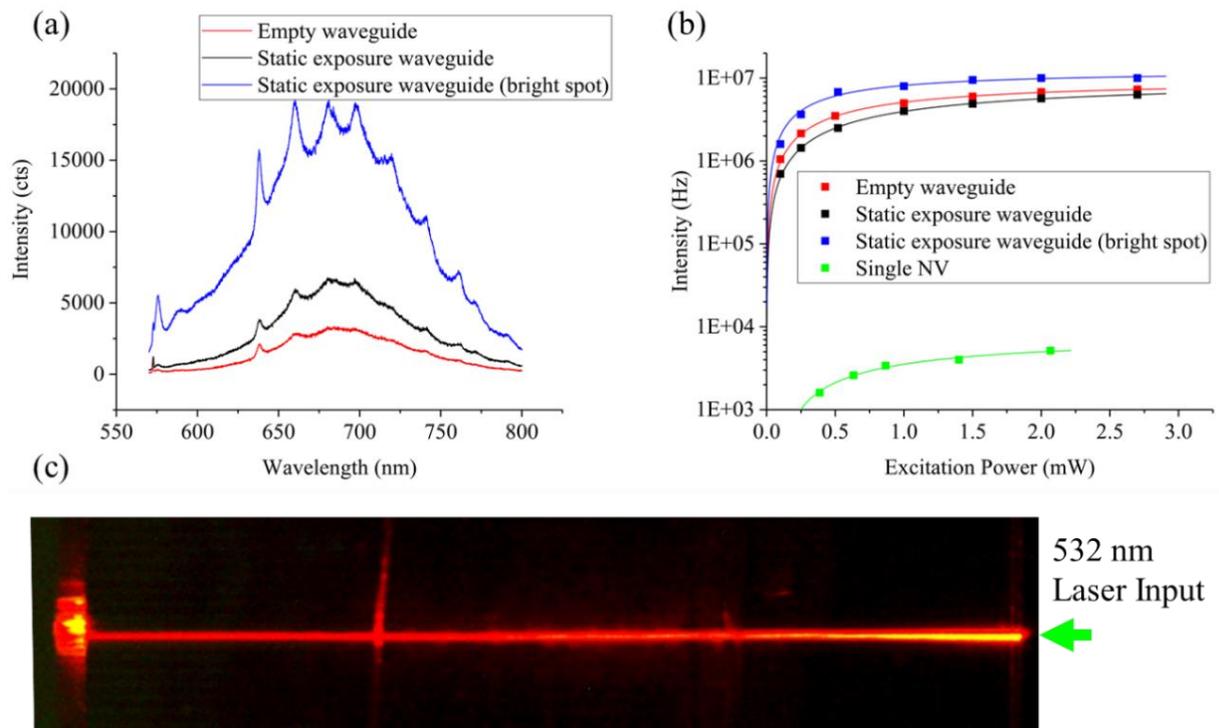

**Figure 11.** (a) PL spectra from 'empty' waveguide, 'static exposure' waveguide, and from a bright spot corresponding to a static exposure. (b) Power dependent PL of 'empty' waveguide, 'static exposure' waveguide, and from a bright spot corresponding to a static exposure. (c) Overhead microscope view of 'static exposure' waveguide. The sample was excited with a green laser end-fire coupled into the waveguide from the right, while the light scattering from the waveguide was filtered using a 532 nm notch filter from the top of the sample and recorded on a CCD.

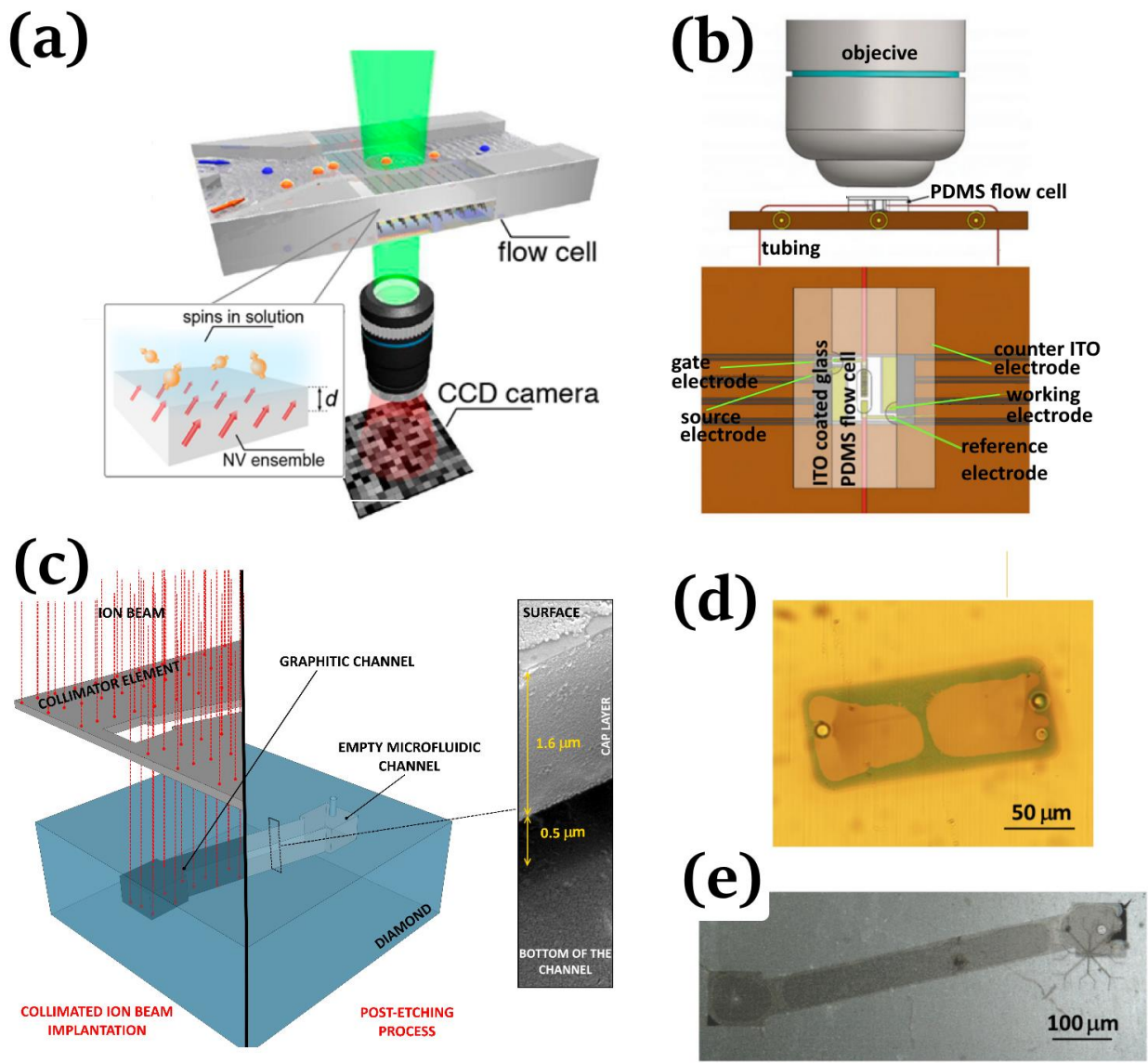

**Figure 12.** (a) and (b) Examples of microfluidic sensors based on quantum detection schemes, adapted from Ziem *et al.*[187] and Krečmarová *et al.*[188], respectively. (c) schematic of the monolithic fabrication of a microfluidic channel by means of ion beam lithography, adapted from Picollo *et al.*[195] and optical micrograph of the obtained structures (d) adapted from Strack *et al.*[194] and (e) adapted from Picollo *et al.*[195].

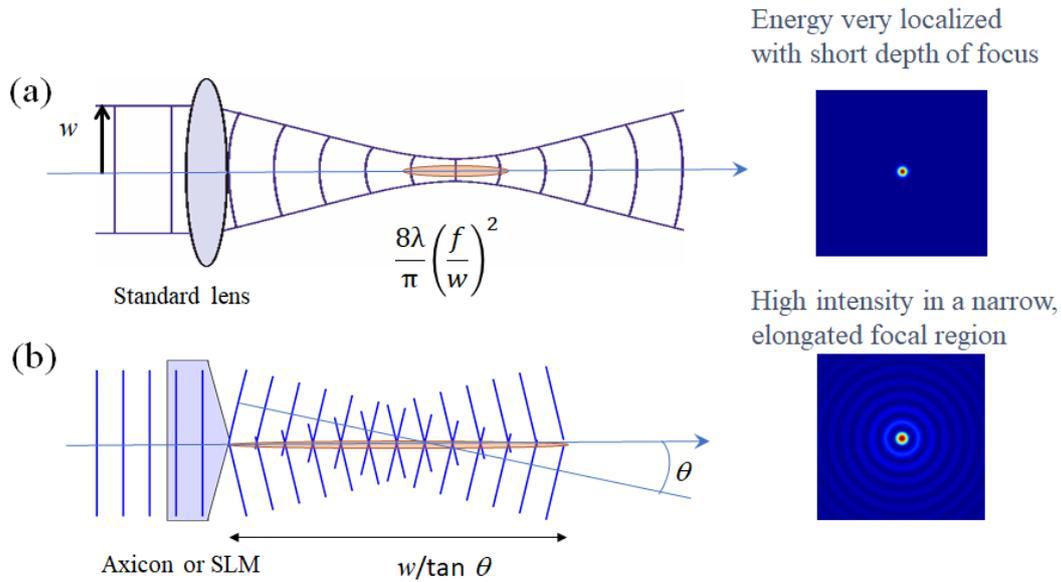

**Figure 13.** (a) Focusing a Gaussian beam with a standard lens: the focal volume is determined by the transverse spot size and the depth of focus, typically leading to a high degree of spatial energy confinement in both radial and axial directions. (b) Focusing with axicon or SLM to produce a Bessel beam, leading to a high intensity in a narrow but axially elongated focal region.

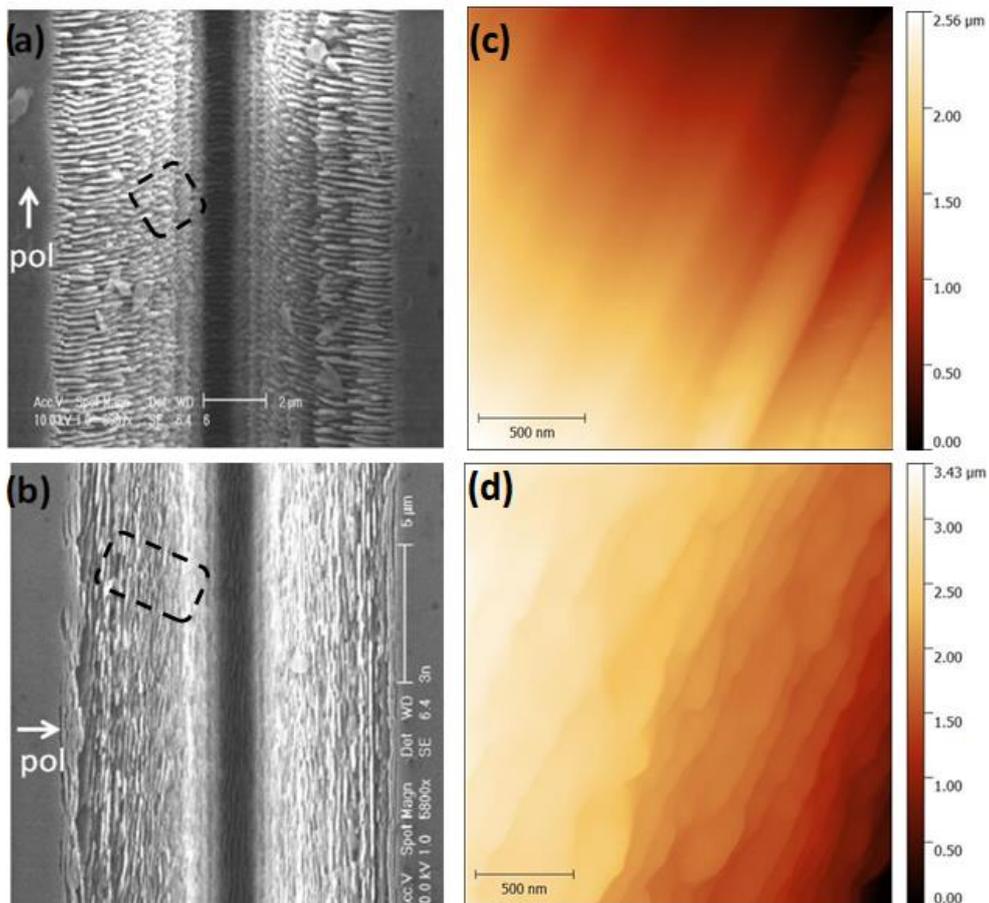

**Figure 14.** Top view SEM images of the microchannels formed inside typical machined trenches for two different transverse writing directions; respectively parallel (a) and orthogonal (b) to the laser beam polarization. In c) and d) AFM tapping-amplitude images of the labelled (dashed) regions in the SEM images[193].

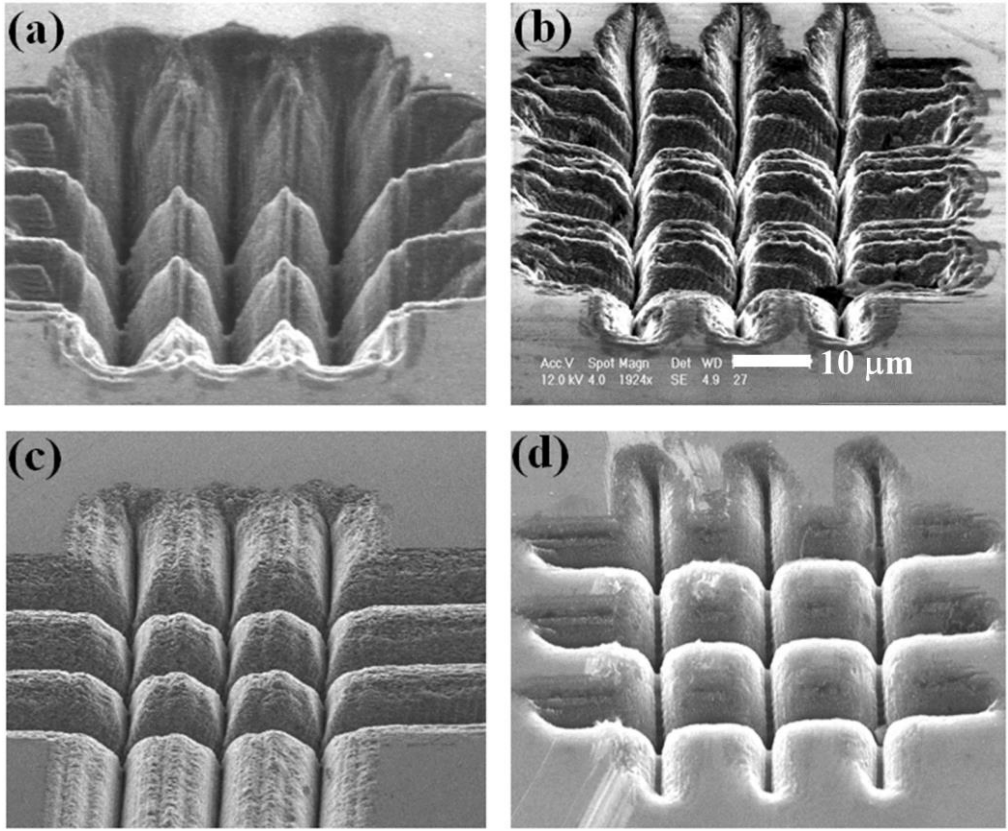

**Figure 15.** SEM images of pillar-like microstructures arrays written on a diamond surface by a 200 fs pulsed Bessel beam in transverse writing configuration and for different cone angles and pulse energies (**a**) $\theta = 9°$, $E = 5.4$ μJ, (**b**) $\theta = 12°$, $E = 4.5$ μJ, (**c**) $\theta = 20°$, $E = 3.5$ μJ and (**d**) $\theta = 24°$, $E = 3$ μJ. Distance between the writing trajectories is 15 μm. The 10 μm scale bar shown in (**b**) is the same for all images[199].

**Keyword** Diamond Photonics

Shane M. Eaton,†, J. P. Hadden†, Vibhav Bharadwaj†, Jacopo Forneris, Federico Picollo, Federico Bosia, Belen Sotillo, Argyro N. Giakoumaki, Ottavia Jedrkiewicz, Andrea Chiappini, Maurizio Ferrari, Roberto Osellame, Paul E. Barclay, Paolo Olivero, Roberta Ramponi*

**Quantum micro-nano devices fabricated in diamond by femtosecond laser and ion irradiation**